\documentclass[sigplan,screen,sigconf]{acmart}

\setcopyright{rightsretained}
\acmDOI{}
\acmISBN{}
\acmConference[LATTE '25]{5th Workshop on Languages, Tools, and Techniques for Accelerator Design}{March 30, 2025}{Rotterdam, Netherlands}

\author{Benedict Short}
\affiliation{
	\institution{Imperial College London}
	\country{UK}
}
\author{Ian McInerney}
\affiliation{
	\institution{Imperial College London}
	\country{UK}
}
\author{John Wickerson}
\affiliation{
	\institution{Imperial College London}
	\country{UK}
}

\begin{document}
	
	\title{Hardware.jl --- An MLIR-based Julia HLS Flow (Work in Progress)}
	
	\begin{abstract}
		Co-developing scientific algorithms and hardware accelerators requires domain-specific knowledge and large engineering resources. This leads to a slow development pace and high project complexity, which creates a barrier to entry that is too high for the majority of developers to overcome. We are developing a reusable end-to-end compiler toolchain for the Julia language entirely built on permissively-licensed open-source projects. This unifies accelerator and algorithm development by automatically synthesising Julia source code into high-performance Verilog.
	\end{abstract}
	\maketitle
	\section{Introduction}
	Rising throughput and latency demands from scientific algorithms mean that they can no longer rely on advancements in computer architecture or fabrication processes for uplifts in performance. Instead, they must take advantage of changes across the stack by reallocating silicon budget to custom accelerators in both FPGA and ASIC designs \cite{ChaNei:20}. Resource-constrained environments, such as embedded real-time control systems and signal processing algorithms, have already adopted this approach to develop next-generation electric motors and software-defined radios. This, however, comes at the cost of developing and maintaining two separate implementations at different layers of abstraction known as the "two-language" problem \cite{BigMcI:22} --- algorithms are developed using mathematically-friendly languages like Julia or Matlab and the accelerators are specified at the RTL level using Verilog or VHDL. Unifying the development process with HLS tools results in faster prototyping, lower development costs and makes this technology accessible to engineers and researchers beyond the hardware domain. 
	
	\subsection{Toolchain Objectives}
	The objectives of our toolchain are outlined below:
	\begin{itemize}
		\item Facilitate quick end-to-end accelerator design
		\item Be reusable and easy-to-maintain
		\item Enable direct hardware synthesis from pure Julia source code
		\item Support common hardware synthesis tools (extensibility)
	\end{itemize}
	
	\subsection{Why Julia?}
	Julia \cite{BezEde:17} is an open-source, high-performance, dynamically typed language built on top of the LLVM framework \cite{LatAdv:04} aiming to solve the "two-language" problem \cite{BigMcI:22}. Its popularity in the scientific community comes from its intuitive syntax, which closely resembles mathematical notation. Julia's extensible compiler infrastructure has promoted the development of cross-compilation frameworks for accelerators using different programming paradigms, such as GPUs \cite{BesChr:19}, TPUs \cite{KenEll:18} and GraphCore IPUs \cite{Gio:23}. Another key advantage is that the ecosystem mainly supports `pure Julia packages' making it more efficient to compile entire libraries to different targets, unlike popular alternatives such as Python.
	\subsection{What about existing toolchains?}
	Existing HLS toolchains for high-level languages, such as Matlab \cite{MathWorksHLS}, are proprietary and generate poorly optimised designs \cite{CurFio:23}, while other HLS tools operate at low levels of abstraction not used for algorithm development (e.g synthesising C/C++, Domain-Specific Languages and RTL equivalents) \cite{CanJon:13, FerCas:21, JosGue:20, YeHao:22, DemLan:22}.
	
	In 2022, Biggs et al.~\cite{BigMcI:22} proposed a solution that successfully compiled control-flow Julia programs into dynamically scheduled VHDL, but noted that their tool suffered from a lack of hardware-specific compiler optimisations and that future work would require more program information to be extracted to support memory and vector operations. Our work now solves this by implementing an MLIR-based workflow to leverage reusable compiler infrastructure and more powerful intermediate representations.
	
	At JuliaCon 2024, Lounes posited the possibility of using Julia as an MLIR front end to integrate with existing statically scheduled HLS infrastructure \cite{Lou:24}, but we take this idea a step further and implement a fully reusable HLS pipeline native to Julia that leverages both static and dynamic scheduling. 
	
	\section{MLIR-based flow}
	We use MLIR \cite{LatMeh:20} to address fragmented HLS pipelines and reduce the cost of developing a new domain-specific compiler by reusing established dialects and interfaces. Standardised infrastructure also makes our solution easy to maintain and allows for innovation at higher levels of abstraction to avoid `reinventing the wheel'. 
	\subsection{Extracting MLIR}
	Julia's nominative, dynamic and parametric type system enables quick prototyping and language-level polymorphism, but the inflexibility of statically compiled accelerators presents an inherent challenge for this project. Standard MLIR dialects allow us to offload this challenge to the front end and enforce type-stable IR code to avoid exponentially increasing the overall design size. Lattner et al. also describe premature lowering as "the root of all evil" \cite{LatMeh:20}, emphasising the importance of extracting high-level dialects.
	
	\subsection{Converting MLIR to Verilog}
	We leverage the open-source CIRCT \cite{CIRCT} framework to deterministically generate syntactically correct Verilog and keep our end-to-end compilation stack within the MLIR ecosystem. CIRCT directly accepts a selection of standard MLIR dialects and provides a library of hardware-specific dialects and passes, allowing us to develop a fully customisable HLS back end that not only supports both dynamic and static scheduling, but also targets a wide range of existing synthesis tools. We also considered using other MLIR-based back ends such as Dynamatic \cite{JosGue:20} and ScaleHLS \cite{YeHao:22}, but they are too restrictive to be used in a stand-alone fashion as they only support either dynamic or static scheduling respectively. Instead, they could be integrated as part of a larger back end that combines these tools in a hybrid manner, similar to the Dynamic and Static Scheduling (DASS) toolchain proposed by Cheng et al. \cite{CheJos:21}.
	
	\section{System Architecture}
\begin{figure}[h]
		\centering
		\includegraphics[width=\linewidth]{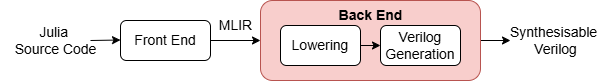}
		\caption{Tool Architecture}
		\label{fig:architecture:high_level_architecture}
	\end{figure}

	The toolchain architecture is inspired by the Bambu HLS framework \cite{FerCas:21}. It is split into three separate components to promote maintainability and modularity, as shown in Fig \ref{fig:architecture:high_level_architecture}.
	
	\subsection{Front End}
	Our front end differs from other tools by mapping Julia constructs to hardware-friendly MLIR instead of focussing entirely on high-level concepts. The \texttt{Core.Compiler} package allows us to create an \texttt{AbstractInterpreter} that uses a custom inlining policy and then lowers source-code into type-stable Julia SSA-form IR. The \texttt{AbstractInterpreter} also allows us to implement method table overlays to integrate directly with Julia's JIT interpreter and support the notion of `world age'. We take advantage of Julia's built-in type inference to support compile-time polymorphism and reuse standard high-level optimisations, such as converting dynamic dispatch into static dispatch. In the majority of cases, this approach enables a one-to-one mapping between the Julia typed IR and its standard MLIR equivalent, similar to the \texttt{Brutus.jl} project \cite{ChuShe:20}. The custom inlining policy allows us to overcome the inherent lack of \texttt{call} and \texttt{invoke} to generate MLIR that is suitable for hardware synthesis. The end of the pipeline applies lifting passes to raise source code to higher levels of abstraction.
	
	\subsection{Back End}
	Currently, the back end supports both static and dynamic scheduling by incrementally lowering standard MLIR to the \texttt{handshake} and \texttt{calyx} dialects. This is then further lowered to the \texttt{SystemVerilog} dialects and used to generate synthesisable Verilog. The current architecture and lowering routes are inspired by the HLS tool provided with the CIRCT library. Communication with the back end is implemented using a standardised interface, making it a seamless process to integrate additional back ends into our toolchain.
	
	\section{The roadmap}
	This project is currently under continuous development and will be made open-source with permissive licensing. The planned future steps are as follows.
	
	\subsection{Front End}
	The first aim will be to incorporate a larger subset of the Julia language to synthesise a wide range of existing Julia packages and libraries. We will also create a custom dialect to avoid a monolithic architecture and produce a standard interface for complex language constructs, such as arrays, that cannot be trivially mapped to their MLIR equivalents. The end goal is to take advantage of statically known program information to make unsupported constructs synthesisable, for example, exploiting knowledge that dynamic arrays will never be larger than a given size to convert them into static arrays. Another aim is to raise Julia directly to Polyhedral MLIR to avoid premature lowering and allow for greater optimisation opportunities, similar to the \texttt{Polygeist} project for C/C++ \cite{MosChe:21}.
	
	\subsection{Back End}
	We will write a wrapper for the CIRCT library to bring the entire HLS back end to Julia. This will provide a platform for HLS research that will allow tools to be prototyped without needing to build LLVM, CIRCT or C++ projects and consequently speed up development time. In future, this could be extended further by implementing advanced identification of static and dynamic islands \cite{CheJos:21} to produce highly optimised HLS designs.
	
	\subsection{Evaluation}
	The compiler toolchain will be evaluated against a set of standard benchmarks to determine performance and design correctness, similar to existing Julia GPU compilation frameworks. The benchmarks shipped with Julia are very limited, and so a wide range of test programs need to be written to provide a quantitative evaluation of the performance of this compiler stack. The tool will also be evaluated against the high-level ideology to ensure that it is user-friendly and integrates well into the existing Julia ecosystem.
	
	\subsection{Testing the compilation stack}
	First, we plan to take advantage of CIRCT's formal verification tooling and built-in logical equivalence checker to verify the correctness of the lowering passes. Concerns have been raised about the reliability of HLS toolchains \cite{HerWic:21} and we will overcome this by using a simple Julia fuzzer to automatically generate synthesisable designs and check for equivalence.
	
	\section{Long-term Vision}
	The AMD Vitis HLS tool supports source-level testbench cosimulation to test the correctness of generated designs against a golden model. Integrating this into the \texttt{Hardware.jl} compiler toolchain would significantly increase the implicit trust that end-users have in the tool. The \texttt{ESI} dialect within CIRCT provides the \texttt{cosim} endpoint, which could be used as a starting point for this work.
	
	Alternative use cases of CIRCT's formal verification framework would be greatly beneficial to this project and would allow the HLS developers to provide guarantees on design correctness or performance. This would also be entirely reusable across all CIRCT projects.
	
	\newpage
	\bibliography{bib}
\end{document}